\documentclass[sn-apa]{sn-jnl}

\usepackage{graphicx}%
\usepackage{multirow}%
\usepackage{amsmath,amssymb,amsfonts}%
\usepackage{amsthm}%
\usepackage{mathrsfs}%
\usepackage[title]{appendix}%
\usepackage{xcolor}%
\usepackage{textcomp}%
\usepackage{manyfoot}%
\usepackage{booktabs}%
\usepackage{algorithm}%
\usepackage{algorithmicx}%
\usepackage{algpseudocode}%
\usepackage{listings}%
\usepackage{appendix}
\usepackage[T1]{fontenc}
\usepackage{pifont}
\usepackage{siunitx}

\newcommand{\I}[1]{\mathbf{1}\!\left\{#1\right\}}

\newcommand{\on}{\ding{51}}      
\newcommand{\off}{\textendash}

\theoremstyle{thmstyleone}%
\theoremstyle{thmstyletwo}%

\theoremstyle{thmstylethree}%

\newif\ifarxiv
\arxivtrue

\begin{document}

\title{STAMP: A shot-type-aware areal multilevel Poisson model for league-wide comparison of basketball shot charts}

\ifarxiv
\author[1]{\fnm{Kazuhiro} \sur{Yamada}}\email{yamada.kazuhiro.t5@s.mail.nagoya-u.ac.jp}
\author*[1,2]{\fnm{Keisuke} \sur{Fujii}}\email{fujii@i.nagoya-u.ac.jp}

\affil[1]{\orgdiv{Graduate School of Informatics}, \orgname{Nagoya University}, \orgaddress{\city{Nagoya}, \country{Japan}}}
\affil[2]{\orgdiv{Center for Advanced Intelligence Project}, \orgname{RIKEN}, \orgaddress{\city{Osaka}, \country{Japan}}}
\else
Anonymous
\fi

\abstract{
Shooting location is a core indicator of offensive style in invasion sports. Existing basketball shot-chart analyses often use spatial information for descriptive visualization, location-based efficiency modeling, or clustering players into shooting archetypes, yet few studies provide a unified framework for fair comparison of shot-type-specific tendencies. We propose the shot-type-aware areal multilevel Poisson (STAMP) model, which jointly models team-level field-goal attempts across predefined court regions, seasons, and shot types using a Poisson likelihood with a possession-based exposure offset. The hierarchical random-effects structure combines team, area, team-area, and team-side random effects with shot-type-specific random slopes for key shot categories. We fit the model using approximate Bayesian inference via the Integrated Nested Laplace Approximation (INLA), enabling efficient analysis of more than $3\times 10^{5}$ shots from two seasons of B.LEAGUE (the men's professional basketball league in Japan). The STAMP model achieves better out-of-sample predictive performance than simpler baselines, yielding interpretable relative-rate maps and left-right bias summaries. Case studies illustrate how the model reveals team-specific spatial tendencies for comparative analysis, and we discuss its limitations and potential extensions.
}





\maketitle

\section{Introduction}
\label{sec:introduction}
In invasion games, shooting is one of the most crucial actions. Beyond its practical value for performance evaluation and scouting, shot data have also become a testbed for methodological innovation. This type of work is especially prevalent in soccer and basketball; however, because their purposes and tasks differ, the modeling strategies adopted in each field often diverge.

In soccer, the relative rarity of goals, which are the most critical event, has caused repeated attention to the value of shots. The most well-known metric for shots in soccer is the Expected Goal (xG), which represents the probability that a given shot results in a goal \citep{anzer2021agoal, lucey2015quality, rathke2017examination, xu2025skor}. Using tracking data (time-series data that continuously record player or ball positions/velocities during a game) and event data (discrete data that record action types and coordinates during a game, such as passes), situation-dependent goal probabilities are modeled and quantified. The models used include logistic regression \citep{eggels2016explaining,alexander2018spatial,marc2020dealing,pardo2020creating}, gradient boosting trees such as XGBoost \citep{chen2016xgboost}\citep{pardo2020creating,cavus2022explainable,mead2023expected,HEWITT2023amachine}, and in some cases (Bayesian) hierarchical models \citep{tureen2022estimated, scholtes2024bayes}. 

By contrast, in basketball, while research attempting to capture shot quality does exist \citep{schmid2025getting,kambhamettu2024quantifying}, it is relatively scarce, partly because player posture data, although critical, are not widely available. Instead, work that visualizes or quantifies shooting efficiency or tendencies is more common \citep{yamada2025two}. Previous studies \citep{reich2006spatial,miller2014factorized,scrucca2025model,wong2023joint,cao2025how,zuccolotto2023spatial,zuccolotto2021spatial,ho2025abayesian} have centered on modeling spatial information using shot charts, which record the locations of shots taken. This literature can broadly be divided into two strands: studies focusing on shooting efficiency by location and studies focusing on spatial shooting tendencies. The former mainly addresses the question of where shots should be taken, with an eye toward applications in optimizing shot strategy \citep{jiao2021bayesian,Sandholtz2020measuring,ehrlich2024estimating}. The latter is primarily concerned with understanding players’ or teams’ spatial preferences, sometimes as a basis for clustering them into archetypes. For this second strand, the modeling approach most frequently applied to shot charts has been the Log-Gaussian Cox Process (LGCP; \cite{moller1998log}) \citep{miller2014factorized,Sandholtz2020measuring,hu2021bayesian,yin2023analysis,cao2025what}.  

LGCP models the log-intensity over continuous space as a Gaussian process, typically estimated via a discretized, sparsity-inducing representation. However, when marks, that is, categorical labels associated with each event (e.g., shot type: jump shot, layup, etc.), are introduced, the continuous field can absorb variation that should instead be attributed to those categories. For comparative inference across players or teams, one natural approach is to introduce random effects indexed by grouping factors (such as team or player) and to impose centering or sum-to-zero constraints so that these effects are identifiable and interpretable as deviations from a common baseline. These additions, however, increase coupling among parameters, reduce sparsity, and worsen the conditioning of the Hessian, leading to numerical instability and higher computational cost. In practice, LGCPs can therefore be cumbersome for fair comparisons between players or teams, especially when mark information is required.

Accordingly, we introduce the shot-type-aware areal multilevel Poisson (STAMP) model, a Poisson generalized linear mixed model (GLMM) for shot counts over predefined court regions. By incorporating possession counts as an offset term and random effects for team, area, team$\times$area, team$\times$side and shot-type-dependent team$\times$area slopes, STAMP yields exposure-adjusted comparisons across teams. Mathematically, the specification is a latent Gaussian model, which enables fast Bayesian inference via the Integrated Nested Laplace Approximation (INLA; \cite{rue2009approximate}). This structure allows us to fit league-scale, shot-type-aware spatial models in minutes; thus, the method is practical for routine performance analysis.

Similarly, shot counts over predefined areas have been modeled using a Poisson GLMM \citep{wong2023joint}. Their goal, however, is player categorization while accounting for shot success, whereas our focus is comparative inference for players and teams. In particular, we emphasize team-level shot charts as a primary unit of analysis, which has been less frequently addressed in prior spatial shot-chart work. Furthermore, by using possession counts as an offset, we mitigate concerns regarding comparability based on exposure, a challenge in that study, enabling rate-based comparisons even when exposure levels differ. In parallel, marked spatial point process approaches have been proposed \citep{jiao2021bayesian,yeung2025transformer}. In contrast, our model is more interpretable, as it adopts a multiplicatively separable, log-linear decomposition on the rate scale, and it is more scalable because inference relies on sparse GLMM/INLA machinery rather than Markov Chain Monte Carlo (MCMC) or deep neural network-based models.

Our contributions are threefold: 
\begin{itemize}
  \item \textbf{Model: reusable template for marked spatial count data.}
  We formulate the STAMP model as a latent Gaussian Poisson GLMM for spatially structured count data with multiplicatively separable, log-linear rate components indexed by team, season, area, side, and shot type.
  Shot type plays the role of a categorical mark, with mark-specific random slopes on team-by-area effects; all random effects (team, area, team$\times$area, team$\times$side, and mark-specific slopes) are mean-centered and constrained to sum to zero, ensuring identifiability and interpretable log-rate contrasts.
  While our empirical study focuses on basketball, the same template can be applied to other invasion sports or spatially aggregated event processes defined over pre-specified regions and categorical marks.

  \item \textbf{Inference \& scalability: fast and stable.}
  The STAMP model remains within the latent Gaussian class and is fitted via INLA with Penalized Complexity (PC) priors on variance and correlation hyperparameters \citep{simpson2017penalising}, avoiding MCMC and heavy deep architectures while
  still providing stable approximate Bayesian inference for league-scale data (on the order of $3\times 10^5$ shots) in roughly a minute per model on a standard CPU.

  \item \textbf{Utility: league-wide interpretability and fair comparisons.}
  The model provides interpretable log-rate contrasts, supports exposure-fair team-to-team comparisons by incorporating possession counts as an offset. In the experiments, we demonstrate the predictive validity and practical application in the B.LEAGUE (the men's professional basketball league in Japan).
\end{itemize}

In brief, the STAMP model is a Poisson GLMM for areal shot counts with a possession offset and fixed effects for shot-type, season, and court side. It includes four key random-effects components, two interaction terms (team$\times$area and team$\times$side) and two shot-type-specific team$\times$area slopes, that capture team- and location-specific deviations in shot tendencies within a latent Gaussian INLA framework with PC priors on variance and correlation hyperparameters. Using only priors that pass prior predictive checks, we then systematically evaluate out-of-sample predictive performance under alternative random-effects specifications obtained by switching on and off these four components. Full details of the model specification, prior choices, and experimental design are given in Methods~\ref{sec:methods}. Guided by this specification, our empirical analysis is designed to address three questions. First, we quantify how much each of the four random-effects components contributes to out-of-sample predictive performance by comparing the sixteen candidate structures obtained by switching them on and off (Section~\ref{sec:core_struct}). Second, we investigate what spatial and left-right patterns the fully specified STAMP model reveals at the league and team levels, focusing on representative B.LEAGUE teams (Section~\ref{sec:spatial_patterns}). Third, we examine whether adding further team-level slopes for shot-type preferences yields meaningful gains in predictive performance beyond the core STAMP structure (Section~\ref{sec:additional_slopes}).

\section{Results}
\subsection{Dataset}
\label{sec:data_preproc}
In this study, we use play-by-play data for a total of 2{,}352 games from the 2023--24 and 2024--25 B.LEAGUE seasons.
Play-by-play data are time-stamped logs that record on-court actions such as shots and rebounds, as well as game events such as substitutions. Note that B.LEAGUE play-by-play data is available for viewing on the official website for each game, but it is not provided as open data in a machine-readable format for bulk acquisition for research purposes. This study utilized play-by-play data with shot location provided for research purposes by Data Stadium Inc., the official data supplier for B.LEAGUE.

In the B.LEAGUE, each team plays against an opponent in rounds consisting of one or two consecutive games; during the regular-season, B1 teams play 60 games over 36 rounds, and B2 teams play 60 games over 32 rounds.
Afterwards, the top eight teams in each division advance to a post-season tournament played in a best-of-three (two-win) format.

\begin{figure}
    \centering
    \small
    \includegraphics[width=0.4\linewidth]{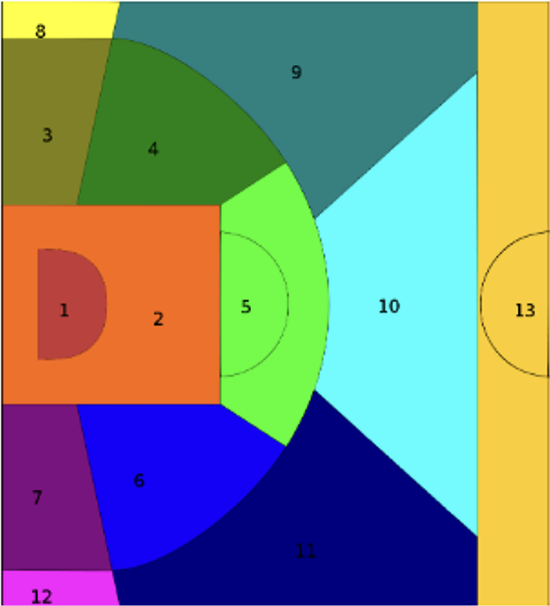}
        \caption{Half-court area divisions; 1: under basket, 2: in the paint, 3: inside right wing, 4: inside right, 5: inside center, 6: inside left, 7: inside left wing, 8: outside right wing, 9: outside right, 10: outside center, 11: outside left, 12: out side leftwing, 13: backcourt.}
    \label{fig:area}
\end{figure}

For modeling, we use only the number of field-goal attempts in each court area for each shot-type, after standardizing the offensive direction, together with the number of possessions for each team.
The shot-type is determined from the action type field in the play-by-play data, and the court area divisions follow the definitions commonly used in the B.LEAGUE (Figure~\ref{fig:area}).
However, left and right symmetric areas are merged into a single area (for example, \textit{outside left} and \textit{outside right} are merged into \textit{outside}).
We also merge several similar shot-type labels to ensure sufficient counts in each category: fadeaway and turnaround jump shot; layup and driving layup; dunk, tip-in, alley-oop, and hook shot; and step-back jump shot and pull-up jump shot, for a total of 6 categories.
The dataset contains 306{,}444 field-goal attempts in total, of which jump shots (denoted \texttt{jump\_shot}) account for 156{,}423 attempts, step-back / pull-up jump shots (denoted \texttt{step\_pull}) account for 50{,}454 attempts, layup group (denoted \texttt{lay\_up}) account for 48{,}550 attempts, floating jump shots (denoted \texttt{floater}) account for 19{,}000 attempts, dunk / tip-in / alley-oop / hook shots (denoted \texttt{rim\_finishes}) acount for 17{,}819 attempts, and fadeaway / turnaround jump shots (denoted \texttt{fade\_turn}) account for 14{,}198 attempts.
The mean number of field-goal attempts per team per season is 4{,}032.16, with a standard deviation of 251.80. Note that each court area is also assigned to a coarse side category (Left, Center, Right) based on Figure~\ref{fig:area}: zones 6, 7, 11, and 12 are treated as Left; zones 1, 2, 5, 10, and 13 as Center; and zones 3, 4, 8, and 9 as Right.

\subsection{Predictive performance of STAMP model} 
\label{sec:core_struct}

We first compare predictive performance across the sixteen 
structures obtained by switching on or off each group of random-effect
components: team$\times$area effects (\texttt{use\_ta}), team$\times$side
effects (\texttt{use\_ts}), shot-type-specific random slopes for jump
shots (\texttt{use\_jump}), and shot-type-specific random slopes for
step-back/pull-up shots (\texttt{use\_step}). All candidates share the
same fixed-effects structure; only the presence
or absence of these random-effect blocks differs (see
Methods~\ref{sec:methods} for details).

\begin{table}[!t]
  \centering
  \caption{Model comparison for core structures. The columns
  \texttt{use\_ta}, \texttt{use\_ts}, \texttt{use\_jump}, and
  \texttt{use\_step} indicate whether each component is included
  (\on) or excluded (\off). The primary criterion is the
  Monte Carlo estimate of the out-of-sample expected log predictive
  density $\widehat{\mathrm{elpd}}_{\text{post}}^{\mathrm{MC}}$, with
  $\mathrm{LPML}_{\text{regular}}$ reported as a secondary,
  fitting-data criterion. Larger values (less negative) indicate
  better predictive performance. All values are computed under the prior
  setting $U_{\mathrm{sd}} = 1.5$ and $U_{\mathrm{slope}} = 1.0$
  described in Section~\ref{sec:prior_spec_ppc}. \textbf{Boldface entries denote the best value among all 16 candidate structures, including those not shown in this table.}}
  \label{tab:model-compare}
  \scriptsize
  \setlength{\tabcolsep}{4pt}
  \sisetup{
    group-separator={,},
    group-minimum-digits=3,
    detect-weight=true,
    detect-inline-weight=math
  }
  \begin{tabular}{
    @{}c c c c
    S[table-format=-6.0]
    S[table-format=-6.0]
    @{}
  }
    \toprule
    \texttt{use\_ta} & \texttt{use\_ts} & \texttt{use\_jump} & \texttt{use\_step} &
    {$\widehat{\mathrm{elpd}}_{\text{post}}^{\mathrm{MC}} \, \uparrow$} & {$\mathrm{LPML}_{\text{reg}} \, \uparrow$} \\
    \midrule
    \off & \off & \off & \off & -6581 & -130617 \\
    \on  & \off & \off & \off & -6422 & -115261 \\
    \off & \off & \on  & \on  & -3499 &  -50141 \\
    \on  & \off & \on  & \on  & -3385 &  \bfseries -37835 \\
    \on  & \on  & \on  & \on  & \bfseries -3383 &  -49967 \\
    \bottomrule
  \end{tabular}

  \vspace{0.6ex}

\end{table}

For each random-effects structure, we fit the STAMP model to regular-season data and evaluate out-of-sample predictive performance on the post-season data. Our primary criterion is the Monte Carlo approximation $\widehat{\mathrm{elpd}}_{\text{post}}^{\mathrm{MC}}$ of the expected log predictive density on post-season possessions, computed by drawing posterior samples of the latent Gaussian field and averaging the log-likelihood contributions cell-wise. As a complementary fitting-data measure, we also report the log pseudo-marginal likelihood on the regular-season data, $\mathrm{LPML}_{\mathrm{reg}}$. Table~\ref{tab:model-compare} summarizes $\widehat{\mathrm{elpd}}_{\text{post}}^{\mathrm{MC}}$ and $\mathrm{LPML}_{\mathrm{reg}}$ for selected core structures under a representative prior configuration.

The simplest baseline that omits all random effects
(\texttt{use\_ta}=$0$, \texttt{use\_ts}=$0$,
\texttt{use\_jump}=$0$, \texttt{use\_step}=$0$) exhibits the worst
predictive performance for post-season data. Adding only
team$\times$area or team$\times$side interactions yields modest
improvements, while the largest gains arise when both interactions and
shot-type-specific slopes are included. The full model with all four
components switched on (\texttt{use\_ta}=$1$, \texttt{use\_ts}=$1$,
\texttt{use\_jump}=$1$, \texttt{use\_step}=$1$) achieves the best
predictive score for post-season, outperforming the baseline by roughly
$\Delta\widehat{\mathrm{elpd}}_{\text{post}}^{\mathrm{MC}}
\approx 3.2\times 10^{3}$ while also attaining the highest
$\mathrm{LPML}_{\mathrm{reg}}$ among the structures with
\texttt{use\_ta}=$1$, \texttt{use\_jump}=$1$, and \texttt{use\_step}=$1$.
However, the gain from including the team$\times$side effects is very
small in out-of-sample terms: the model with
\texttt{use\_ta}=$1$, \texttt{use\_ts}=$0$, \texttt{use\_jump}=$1$,
\texttt{use\_step}=$1$ attains an
$\widehat{\mathrm{elpd}}_{\text{post}}^{\mathrm{MC}}$ only two units
worse than the full model, while achieving a substantially better
$\mathrm{LPML}_{\mathrm{reg}}$. This suggests that most of the predictive
gain comes from the team$\times$area interactions and jump/step
slopes, and that team$\times$side effects provide, at best, a marginal
refinement. Despite its richer random-effect structure, the full model
remains computationally feasible: a single INLA fit requires on the
order of one minute of wall-clock time on our hardware (see
Section~\ref{sec:model_fit_eval}).

To assess the sensitivity of this comparison to the prior
hyperparameters, we repeated the sixteen-model comparison across all
prior configurations that passed the prior predictive checks in
Section~\ref{sec:prior_spec_ppc}. For each random-effects structure, we averaged $\widehat{\mathrm{elpd}}_{\text{post}}^{\mathrm{MC}}$ and
$\mathrm{LPML}_{\mathrm{reg}}$ over these prior settings; the full model remained the best performer in terms of mean $\widehat{\mathrm{elpd}}_{\text{post}}^{\mathrm{MC}}$, and the improvement over the baseline stayed on the order of $3.2\times 10^{3}$. Moreover, for every prior configuration, the highest $\widehat{\mathrm{elpd}}_{\text{post}}^{\mathrm{MC}}$ was always attained by one of the two richest structures,
\texttt{use\_ta}=$1$, \texttt{use\_jump}=$1$, \texttt{use\_step}=$1$
with \texttt{use\_ts} either $0$ or $1$; no sparser specification ever
ranked first. The difference between these two top models was always
small (on the order of a few units in
$\widehat{\mathrm{elpd}}_{\text{post}}^{\mathrm{MC}}$), reinforcing the
view that team$\times$side effects play a secondary role compared with
team$\times$area interactions and shot-type-specific slopes. The ranking of the sixteen structures was also highly stable across prior configurations: Kendall’s $\tau$ rank correlations between the model rankings were above 0.97. Taken together, these results indicate that the qualitative advantage of including
team$\times$area interactions and shot-type-specific random slopes (with or without team$\times$side effects) is robust to reasonable changes in the prior specification.

\subsection{Spatial patterns and interpretation}
\label{sec:spatial_patterns}

To illustrate how the STAMP model with all random effects can be used to interpret spatial tendencies in practice, we first summarize league-wide left-right biases and then present detailed examples for two representative B.LEAGUE teams, Utsunomiya Brex and Nagoya Diamond Dolphins.

\begin{figure}
    \centering
    \small
    \includegraphics[width=0.5\linewidth]{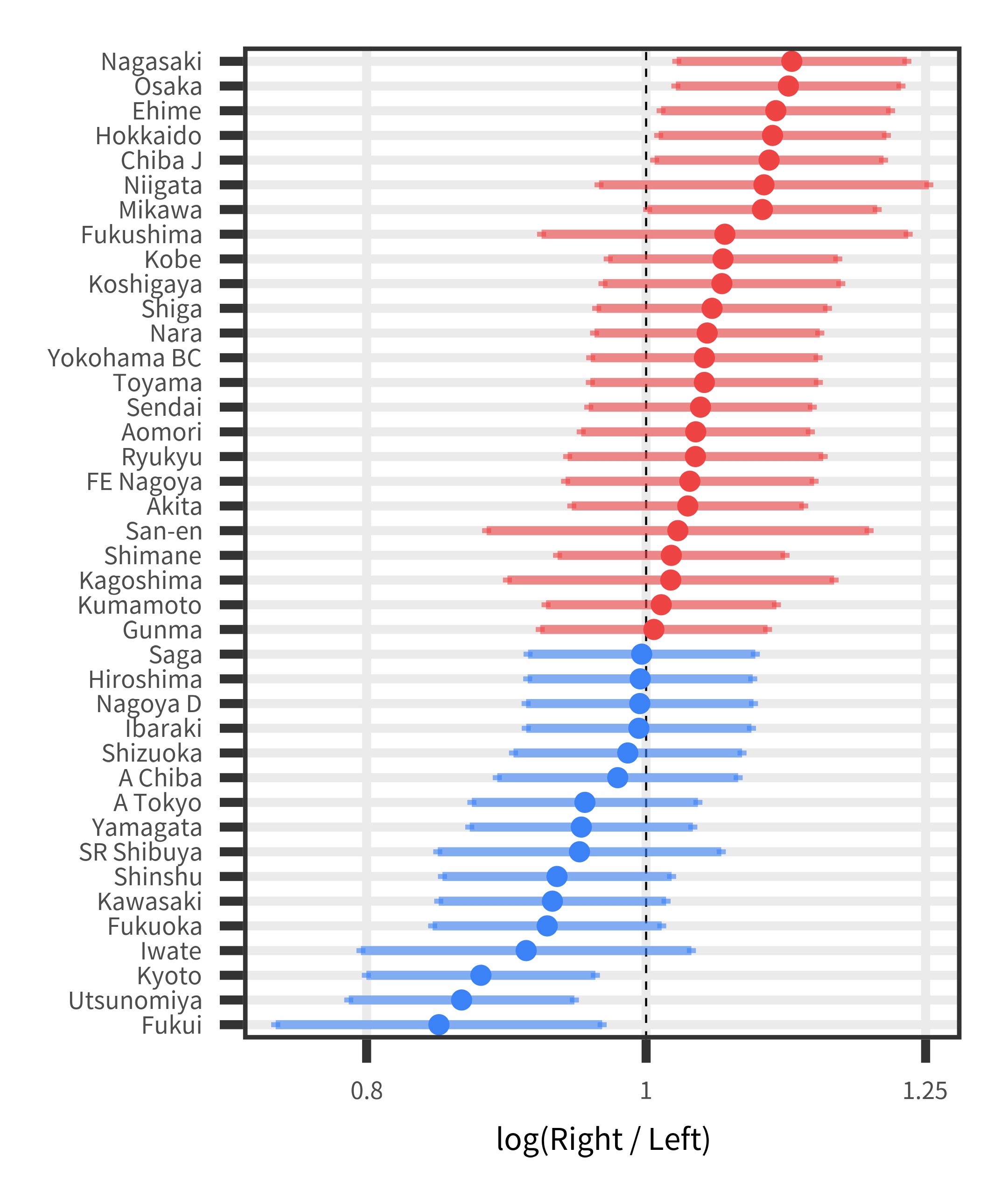}
    \vspace{-12pt}
    \caption{Left-right asymmetry caterpillar plot.
    The horizontal axis shows the logarithm of the ratio of relative occurrence rates between right and left (aggregated for each team using the geometric mean over the season; values closer to the left indicate a bias toward the left side).
    The dashed line indicates no left-right bias, the points represent the estimated values, and the horizontal bars represent the approximate 95\% confidence interval.}
    \label{fig:lr_caterpillar}
\end{figure}

\begin{figure}[!t]
  \centering \includegraphics[width=0.9\linewidth]{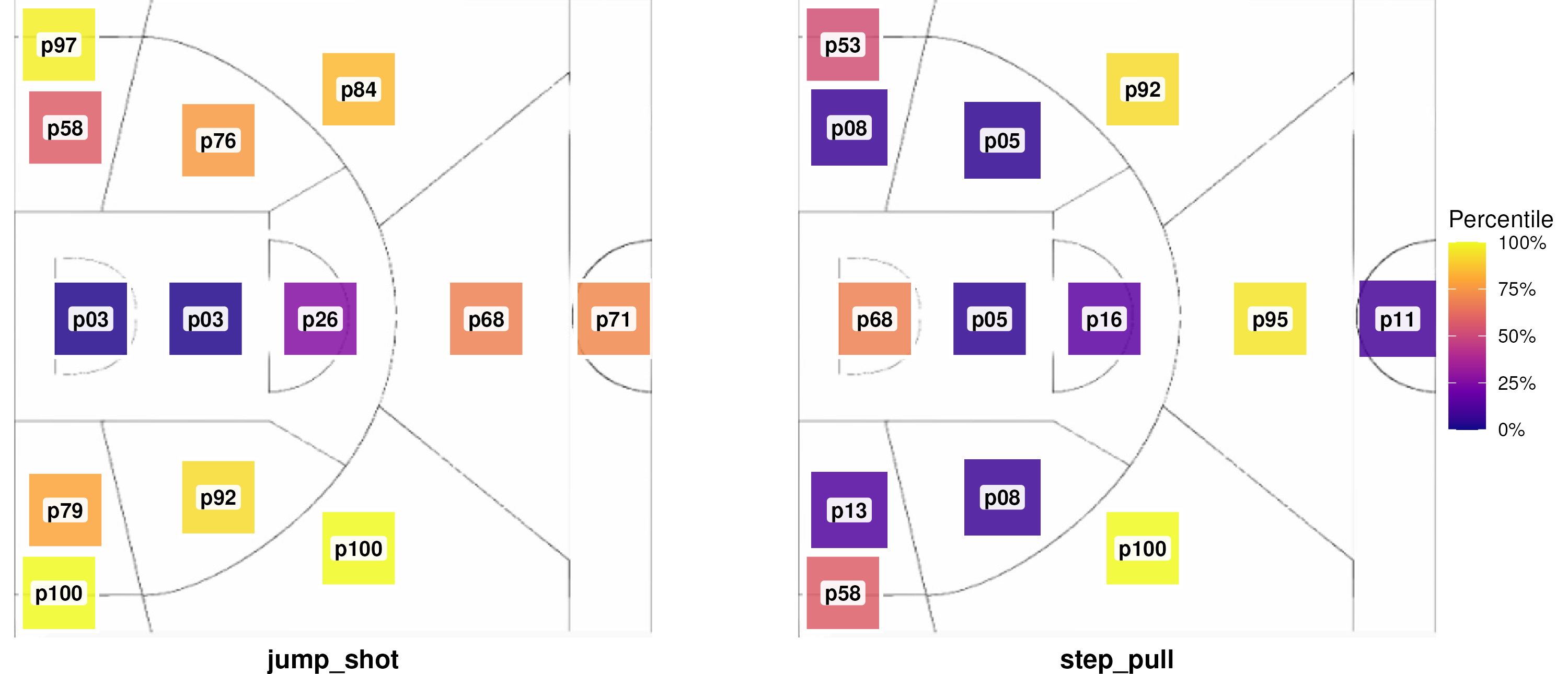}
  \caption{Area-specific relative appearance trends for the Utsunomiya Brex in the 2024-25 season (Left: \texttt{jump\_shot}, Right: \texttt{step\_pull}); side scaling applied. Colors and labels represent team-level percentiles (pXX), with larger values and brighter colors indicating relatively more frequent occurrences.}
  \label{fig:utsunomiya}
\end{figure}

\begin{figure}[!t]
  \centering \includegraphics[width=0.9\linewidth]{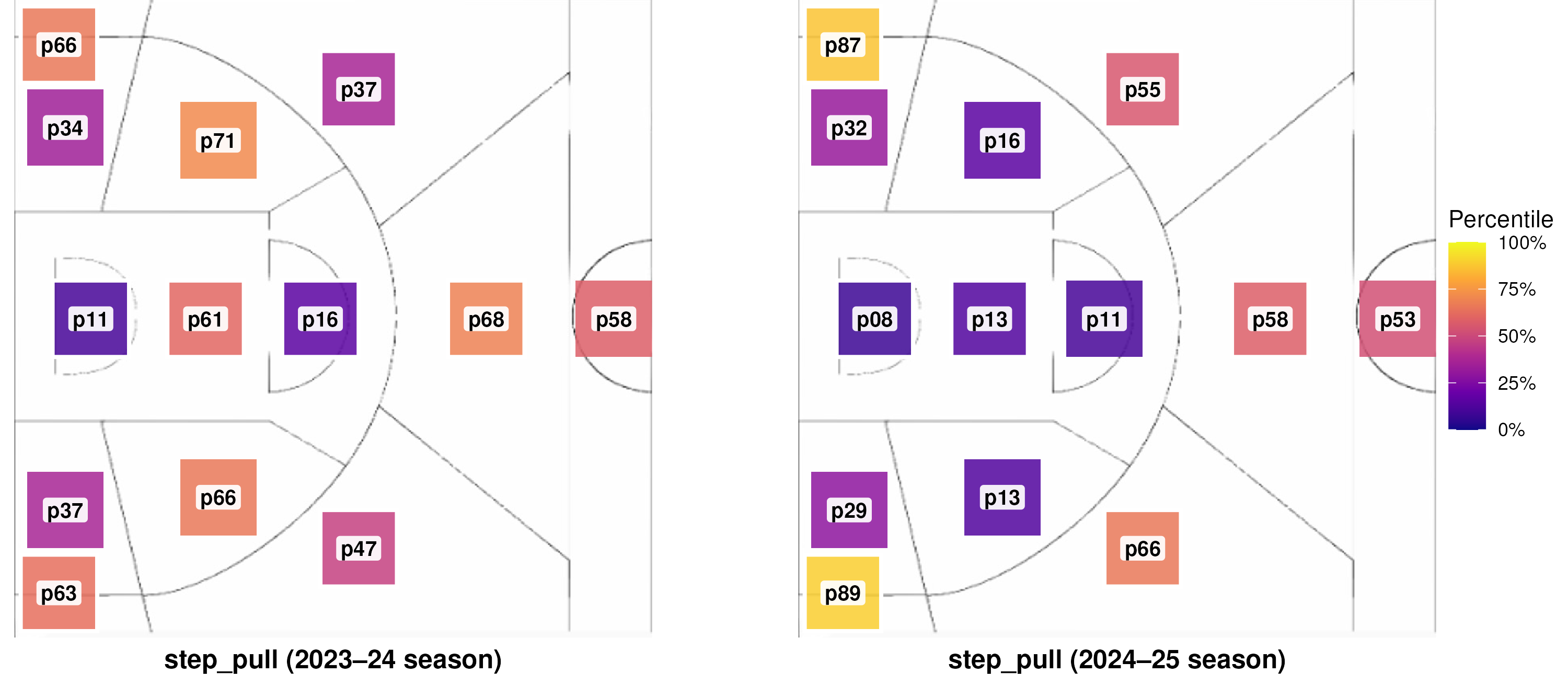}
  \caption{Area-specific relative appearance trends for the Nagoya Diamond Dolphins in the two seasons (left: 2023--24 season, right: 2024--25 season); side scaling applied. Colors and labels represent team-level percentiles (pXX), with larger values and brighter colors indicating relatively more frequent occurrences.}
  \label{fig:nagoya}
\end{figure}

Figure~\ref{fig:lr_caterpillar} shows a caterpillar plot of the
team-by-side effects $z_{ids}$, summarized as the log ratio
$\log(\mathrm{Right}/\mathrm{Left})$ of the posterior multipliers for the right and left sides,
aggregated over seasons via a geometric mean. Values greater than zero
indicate a preference for shots on the right side, whereas negative
values indicate a left-side preference; horizontal bars indicate
approximate $95\%$ credible intervals. Most teams exhibit moderate left-right asymmetries whose intervals overlap
the symmetric case $\log(\mathrm{Right}/\mathrm{Left})=0$, but a few teams show clearly identifiable side-to-side biases. This type of summary is directly interpretable for coaches and analysts: it highlights whether a team tends to favor one side of the court over the other, after adjusting for overall shot volume.

We then focus on two B.LEAGUE teams, Utsunomiya Brex and Nagoya Diamond Dolphins, to provide concrete examples of the spatial patterns captured by the STAMP model. For each team, we combine the estimated shot-type-specific random slopes with the team$\times$side effects and convert the resulting posterior mean rates into percentiles relative to the league-wide distribution for each area. Intuitively, these maps summarize where a team shoots more or less frequently than the league baseline, after accounting for overall shooting volume.

Figure~\ref{fig:utsunomiya} shows side-adjusted shot charts for Utsunomiya Brex in the 2024-25 season, for \texttt{jump\_shot} (left panel) and \texttt{step\_pull} (right panel). Each tile corresponds to one of the predefined court regions, and its color encodes the team-level percentile of the posterior rate among all teams in the same season (brighter colors indicate higher percentiles), while overlaid labels (pXX) report the corresponding percentile in numerical form. By construction, the figure highlights hot and cold regions in terms of relative attempt frequency, rather than raw shot counts, making it easy to identify where Utsunomiya is particularly active or inactive compared with league-wide tendencies. The chart indicates that Utsunomiya is not only a team that attempts many three-point shots, but also shows a relatively strong preference for taking above the break three-pointers when shooting off the dribble. This qualitative agreement with Utsunomiya’s reputation as a three-point-oriented team \citep{onuma2025brex} offers a useful sanity check on the STAMP estimates, suggesting that the model captures salient aspects of their offensive profile.

To illustrate temporal comparisons, Figure~\ref{fig:nagoya} presents
side-adjusted \texttt{step\_pull} charts for Nagoya Diamond Dolphins in the 2023-24 and 2024-25 seasons, displayed side by side. Because the color scale and percentile definition are shared across panels, changes in shading and labels directly reflect shifts in the team’s relative use of each region from season to season. In this way,
STAMP-based maps provide an interpretable league-normalized summary of how spatial tendencies of a team remain stable or evolve.

\subsection{Extended team-level slopes}
\label{sec:additional_slopes}

To examine whether further team-level heterogeneity in shot-type preferences improves predictive performance, we conducted an additional experiment that augments the STAMP model with extra team-level slopes. Fixing $\texttt{use\_ta}=1$, $\texttt{use\_jump}=1$, and $\texttt{use\_step}=1$, we considered all $2\times 2^4 = 32$ model configurations obtained by toggling the inclusion of team-by-side effects ($\texttt{use\_ts}\in\{0,1\}$) and four additional team-level, season-specific slopes for the remaining shot-type categories, which have small spatial variation: \texttt{lay\_up}, \texttt{floater}, \texttt{rim\_finishes}, and \texttt{fade\_turn} (indicators \texttt{use\_lay}, \texttt{use\_float}, \texttt{use\_rim}, \texttt{use\_fade}). All models were fitted under the same PC prior configuration as in the main comparison ($U_{\mathrm{sd}}=1.5$, $U_{\mathrm{slope}}=1.0$;
Section~\ref{sec:prior_spec_ppc}) and evaluated using $\widehat{\mathrm{elpd}}_{\text{post}}^{\mathrm{MC}}$ and $\mathrm{LPML}_{\mathrm{reg}}$. Selected results are summarized in Table~\ref{tab:team-slope-top10}, which lists the top 10 extended-slope configurations ranked by $\widehat{\mathrm{elpd}}_{\text{post}}^{\mathrm{MC}}$.

\begin{table}[!t]
  \centering
  \caption{Top 10 extended team-level slope configurations, ranked by
  $\widehat{\mathrm{elpd}}_{\text{post}}^{\mathrm{MC}}$ (larger values indicate better
  out-of-sample predictive performance). All models include
  team$\times$area interactions and jump/step team$\times$area$\times$season
  slopes (\texttt{use\_ta}=\on, \texttt{use\_jump}=\on,
  \texttt{use\_step}=\on); the table reports which additional components are
  switched on (\on) or off (\off). The prior scales are fixed at
  $U_{\mathrm{sd}}=1.5$ and $U_{\mathrm{slope}}=1.0$. \textbf{Boldface entries denote the best value among all 32 candidate structures, including those not shown in this table.}}
  \label{tab:team-slope-top10}
  \scriptsize
  \setlength{\tabcolsep}{4pt}
  \sisetup{
    group-separator={,},
    group-minimum-digits=3,
    detect-weight=true,
    detect-inline-weight=math
  }
  \begin{tabular}{
    @{}c c c c c
    S[table-format=-4.0]
    S[table-format=-6.0]
    @{}}
    \toprule
    \texttt{use\_ts} & \texttt{use\_lay} & \texttt{use\_float} &
    \texttt{use\_rim} & \texttt{use\_fade} &
    {$\widehat{\mathrm{elpd}}_{\text{post}}^{\mathrm{MC}} \, \uparrow$} &
    {$\mathrm{LPML}_{\text{reg}} \, \uparrow$} \\
    \midrule
    \on  & \on  & \on  & \off & \on  &  \bfseries -3284 & -510493 \\
    \on  & \on  & \off & \on  & \on  & -3308 & -355010 \\
    \on  & \on  & \off & \off & \on  & -3341 &  -73155 \\
    \off & \on  & \on  & \off & \on  & -3347 &  -37184 \\
    \off & \on  & \off & \off & \on  & -3347 &  -37404 \\
    \on  & \on  & \on  & \on  & \off & -3347 & -50299 \\
    \off & \on  & \on  & \on  & \on  & -3348 &  -37297 \\
    \off & \off & \on  & \on  & \on  & -3349 &  -37232 \\
    \off & \on  & \on  & \on  & \off & -3349 &  -37453 \\
    \off & \off & \on  & \off & \on  & -3352 &  \bfseries -36806 \\
    \bottomrule
  \end{tabular}
\end{table}

As shown in Table~\ref{tab:team-slope-top10}, some highly flexible specifications that turn on most of the additional slopes achieve slightly larger $\widehat{\mathrm{elpd}}_{\text{post}}^{\mathrm{MC}}$ than the original full STAMP model, but at the cost of extremely poor $\mathrm{LPML}_{\mathrm{reg}}$ (e.g., on the order of $-5\times 10^{5}$), suggesting overfitting and unstable fit to the regular-season data. Focusing on configurations that simultaneously perform well on both criteria, the best trade-off was obtained by the model with $\texttt{use\_ta}=1$, $\texttt{use\_ts}=0$, $\texttt{use\_jump}=1$, $\texttt{use\_step}=1$, and additional team-level slopes only for \texttt{floater} and \texttt{fade\_turn} (\texttt{use\_lay}=\texttt{use\_rim}=0, \texttt{use\_float}=\texttt{use\_fade}=1; Table~\ref{tab:team-slope-top10}). This configuration improved the postseason criterion to
$\widehat{\mathrm{elpd}}_{\text{post}}^{\mathrm{MC}}\approx -3.35\times 10^{3}$,
about $+3.0\times 10^{1}$ relative to the original full model
($\texttt{use\_ta}=\texttt{use\_ts}=\texttt{use\_jump}=\texttt{use\_step}=1$),
and simultaneously increased
$\mathrm{LPML}_{\mathrm{reg}}$ by roughly $1.0\times 10^{3}$.

Overall, these results indicate that allowing additional team-level slopes for a subset of shot types can yield incremental gains, but the marginal improvement in predictive performance over the original STAMP specification is modest compared to the gains already obtained from including team$\times$area interactions and shot-type-specific jump/step slopes. For parsimony and interpretability, we therefore adopt the full STAMP model without the extra team-level slopes on \texttt{lay\_up}, \texttt{floater}, \texttt{rim\_finishes}, and \texttt{fade\_turn} as our primary specification, and regard this extended-slope analysis as a robustness check. In applications where the main objective is to obtain a concise, stable description of team-level spatial tendencies, such additional slopes may not justify the extra model complexity, whereas they can be selectively introduced when one wishes to sharpen relative comparisons for a particular shot-type category of interest.

\section{Discussion}
\label{sec:discussion}

\subsection{Summary and practical implications}
\label{sec:disc_summary}
The STAMP model extends conventional shot-chart analyses by jointly modelling team-level field-goal attempts across predefined court regions, seasons, and shot types, while remaining computationally feasible for league-scale datasets. The resulting relative-rate maps and left–right bias summaries can provide coaches and analysts with interpretable diagnostics of where and how a team tends to generate shots relative to the rest of the league, as illustrated by the examples for Utsunomiya Brex and Nagoya Diamond Dolphins (Section~\ref{sec:spatial_patterns}).

Beyond the case studies in this paper, we believe that the STAMP model has the potential to serve as a routine monitoring and scouting tool. Because the model only requires play-by-play-level shot records and possession counts, it can be refit as new games accumulate and embedded into existing analytics dashboards without additional tracking infrastructure. The per-possession rate formulation facilitates comparisons fair across teams with different paces, while the percentile-based maps and left-right bias summaries can be overlaid on conventional efficiency statistics (e.g., field-goal percentage by region) to jointly assess ``how often'' and ``how well'' teams shoot from different locations. In practice, such a tool could help coaching staff quickly identify stylistic tendencies and potential blind spots.

\subsection{Methodological implications and robustness}
\label{sec:disc_method}
Methodologically, the STAMP model is specified as a Poisson linear mixed model with Gaussian random effects, which places it squarely within the class of latent Gaussian models targeted by INLA. Working with predefined court regions further helps keep the dimensionality of the latent field manageable and the resulting components interpretable, enabling fitting and inference at league scale without resorting to MCMC or heavy architectures.

Our systematic comparison of structures and prior configurations suggests that the main substantive gains in predictive performance arise from including team$\times$area effects together with shot-type-specific slopes for jump shots and step-back/pull-up jump shots (Section~\ref{sec:core_struct}). This conclusion was robust
across all PC prior scale choices that passed the prior predictive checks, with the richest models consistently ranking at or near the top. The extended experiment with additional team-level slopes for \texttt{lay\_up}, \texttt{floater}, \texttt{rim\_finishes}, and \texttt{fade\_turn} showed only modest incremental improvements in $\widehat{\mathrm{elpd}}_{\text{post}}^{\mathrm{MC}}$ relative to the original STAMP specification. These findings suggest that, unless a specific application calls for detailed relative comparisons within those particular shot-type categories, the simpler slope structure appears to retain most of the predictive and explanatory power while keeping the model more parsimonious. In a similar spirit, side effects are handled in a deliberately simple way in the present specification, through additive team$\times$side interactions only. If an analysis required more detailed side-specific behaviour, the same framework could be extended to allow shot-type-dependent team$\times$side slopes or other side-varying components, at the cost of additional complexity and data requirements.

From a modeling perspective, the STAMP model can be viewed as a pragmatic compromise between fully continuous (marked) point-process formulations (e.g., LGCP) and purely descriptive
shot-chart summaries. By aggregating shots to predefined court regions and modelling region-level attempt rates with a log-linear hierarchical structure, the STAMP model no longer tracks the exact $(x,y)$ location of each shot. Consequently, it cannot represent very fine-grained within-region preferences (for example, a tendency to favor certain spots in the paint). In return, the model works with a much lower-dimensional latent field, enjoys far more replication per parameter, and can be easier to fit and interpret at the league scale. Combined with the partial pooling structure across teams and seasons, this often makes the STAMP model more data-efficient than continuous-space LGCP approaches when applied at the league scale.

The choice of predefined regions is also motivated by practice.
We adopt a predefined partition of the half-court that closely matches the zone layout used in official B.League shot charts, so that the regions remain coarse enough to have a clear tactical interpretation while still being fine enough to reveal substantial differences in shot selection. In applied analysis and scouting, shot charts are often summarized at this regional resolution, so working with these predefined areas preserves practical interpretability while enabling efficient partial pooling.

\subsection{Limitations and future directions}
\label{sec:disc_limits}

This study has several limitations regarding scope and
generalizability. All analyses are based on two seasons of B.LEAGUE play and are carried out at the team level within a single men’s league. We do not examine longer historical windows, women’s competitions, lower divisions, or multi-league settings, nor do we fit player-level models. Consequently, it remains an open question how well the present STAMP specification and its fitted spatial patterns would transfer to other contexts, and applying the model elsewhere would likely require recalibrating priors and possibly adapting random effects.

Second, the current work focuses exclusively on shot frequency rather than shot outcomes or efficiency. We do not model made/missed outcomes, effective field-goal percentage, or expected points, so the STAMP model as specified here is best viewed as a tool for comparing where teams tend to generate shots, not how efficient they are from those locations.

Nevertheless, the same hierarchical framework can be extended to jointly model shot volume and outcomes. A natural next step would be to combine the current Poisson model for attempt counts with an additional binomial (or logistic) submodel for made/missed outcomes, thereby defining both an expected attempt rate (xR) and expected points (xP) at the team-region-shot-type level. Such an extension would retain the interpretability of the present STAMP model specification, while offering a unified view of spatial shooting tendencies and scoring efficiency.

Third, by construction, the STAMP model is an areal model: shots are aggregated to predefined court regions and summarized via a log-linear Poisson hierarchy. As a result, the model cannot resolve fine-grained variation in continuous shot locations within a region, nor does it
explicitly account for defender positions, play type, or other
contextual factors. Such information is averaged within regions or indirectly absorbed into the random effects, which is appropriate for team-level tendency comparisons, but may be limiting for more detailed tactical analysis. If richer mark information, such as play-type labels (e.g., those provided by commercial tagging services), is available, the same areal framework could be extended to incorporate these marks as additional fixed or random effects, or even as alternative responses, thereby enabling complementary models that focus on offensive style or play-context patterns.

An additional modeling choice concerns the shot-type taxonomy used in the present analysis. For modeling convenience and to secure sufficient counts per cell, we manually merged several play-by-play action labels into six broader shot-type categories (e.g., grouping layups and driving layups, or fadeaway and turnaround jump shots). Although these groupings are defensible from a basketball perspective, they are inevitably somewhat ad hoc. If richer contextual information, such as shooting posture, ball trajectory, or player tracking data, were available, the indicator functions that gate the shot-type-specific random slopes could, in principle, be replaced by features or latent representations learned from those data. This would open the door to more data-driven and potentially more nuanced shot-type partitions within the same hierarchical framework, while preserving the overall structure of STAMP.

Beyond team-level applications, extending the STAMP model to player-level comparisons would be a natural next step. A straightforward approach would be to embed additional random effects for players nested within teams, allowing the same regional and shot-type structure to describe within-team variation. For cross-team comparisons, however, it may often be more meaningful to compare players who occupy similar tactical roles (e.g., ball handlers, shooters, bigs). In such settings, one could either fit separate models by role or include role indicators and role-specific random effects so that differences in spatial tendencies are interpreted relative to players with comparable responsibilities and usage, provided that each player accumulates sufficient shot volume. Practically, adding such player- and role-level components also increases the number of hyperparameters and the size of the latent field, which in turn raises the computational cost of INLA. If player comparison is the only objective, a simpler player-level shot-chart model may therefore be a more computationally convenient choice.

More broadly, the STAMP framework is potentially applicable to other invasion sports in which shot actions are spatially structured, such as handball or ice hockey. In contrast, for sports like soccer, where shots are relatively rare and much of the tactical richness lies in the build-up play, it may be more practically useful to adapt the same areal Poisson idea to model passes, entries into dangerous zones, or other intermediate actions using an appropriate exposure offset rather than shots themselves. In this sense, the STAMP model can be regarded as a general template for spatially structured count data that can be repurposed according to the most informative action type in each sport.

\section{Methods}
\label{sec:methods}

\subsection{Proposed model: shot-type-aware areal multilevel Poisson (STAMP) model}
\label{sec:proposed_model}

In this study, for team $i=1,\dots,I$, season $s=1,\dots,S$, area $a=1,\dots,A$, 
side $d\in\{\mathrm{Left},\mathrm{Right},\mathrm{Center}\}$, and shot-type 
$k\in\mathcal{K}$, we model the number of field goal attempts in the cell 
$(i,s,a,d,k)$, denoted by $Y_{isadk}$, using the team$\times$season possession 
count $E_{is}$ as an exposure offset. Throughout, we adopt a shot-terminated definition of possession: a possession ends at each field goal attempt, and an offensive rebound begins a new possession.

\subsubsection{Likelihood and linear predictor}

The likelihood and linear predictor are given by
\begin{align}
  Y_{isadk} &\sim \mathrm{Poisson}(\mu_{isadk}), &
  \log \mu_{isadk} &= \log E_{is} + \eta_{isadk},
\end{align}
where $\eta_{isadk}$ denotes the linear predictor. We decompose
$\eta_{isadk}$ as
\begin{align}
  \eta_{isadk}
  &= \underbrace{\beta_0 + \beta^{(\mathrm{season})}_s
     + \beta^{(\mathrm{side})}_d + \beta^{(\mathrm{shot\_type})}_k}_{\text{fixed effects}}
     \nonumber\\[2pt]
  &\quad +\, \underbrace{u_i}_{\text{team (iid)}}
          + \underbrace{v_a}_{\text{area (iid)}}
     \nonumber\\[2pt]
  &\quad +\, \underbrace{w_{ias}}_{\text{team}\times\text{area (seasonally correlated)}}
          + \underbrace{z_{ids}}_{\text{team}\times\text{side (season-specific)}}
     \nonumber\\[2pt]
  &\quad +\, \sum_{m=1}^{M} \gamma_m(k)\,
         \underbrace{r^{(m)}_{ias}}_{\text{shot-type-specific team}\times\text{area (seasonally correlated)}},
  \label{eq:rs-general}
\end{align}
subject to the sum-to-zero constraints
\begin{align}
\sum_{i=1}^{I} u_i &= 0,\\
\sum_{a=1}^{A} v_a &= 0,\\
\sum_{i=1}^{I}\sum_{a=1}^{A}\sum_{s=1}^{S} w_{ias} &= 0,\\
\sum_{d\in\{\mathrm{Left},\mathrm{Right},\mathrm{Center}\}} z_{ids} &= 0 
  \quad (\forall\, i,s),\\
\sum_{i=1}^{I}\sum_{a=1}^{A}\sum_{s=1}^{S} r^{(m)}_{ias} &= 0 
  \quad (\forall\, m),
\end{align}
where, $E_{is}$ is the number of possessions for team $i$ in season $s$ 
(offset term); $u_i$ is an i.i.d.\ random effect for team $i$; and 
$v_a$ is an i.i.d.\ random effect for area $a$. The terms $w_{ias}$ are 
seasonally correlated team$\times$area effects; for each $(i,a)$ we assume an 
equicorrelation structure 
$\mathrm{corr}(w_{ias}, w_{ias'}) = \rho_{\mathrm{ta}}$ for $s \neq s'$.
For team$\times$side, we use season-specific i.i.d.\ random effects $z_{ids}$ 
with zero-sum constraints $\sum_{d} z_{ids}=0$ for each $(i,s)$, but do not 
impose an explicit temporal correlation across seasons. Since side is a coarse 
factor (Left/Right/Center) and the corresponding counts are relatively 
abundant, we mainly introduce seasonal correlation for the more granular and 
data-sparse team$\times$area and shot-type-specific effects. 

The collection $\{r^{(m)}_{ias}\}_{m=1}^M$ denotes shot-type-dependent random 
slopes for team$\times$area; for each $m$ and $(i,a)$ these are also modeled as 
seasonally correlated across $s$, analogously to $w_{ias}$. The function 
$\gamma_m(k)$ is a known coding function for shot-type $k$. As the simplest 
example, using the indicator function based on the family of shot-type sets 
$\{\mathcal{K}_m\}_{m=1}^{M}$,
\begin{align}
  \gamma_m(k) \;=\; \I{k \in \mathcal{K}_m},
\end{align}
\eqref{eq:rs-general} can be written as
\begin{align}
  \eta_{isadk}
  \;=\; \cdots
  \;+\; \sum_{m=1}^{M} \I{k \in \mathcal{K}_m}\, r^{(m)}_{ias},
\end{align}
and this is the specification adopted in our experiments. Note that each fixed effect is treated as a deviation from a reference category ($\mathcal{N}(0, 100^2)$ is used as a nearly uninformed prior).

Because including a uniform interaction for all shot-types would cause over-parameterization and destabilize cells $(i,s,a,d,k)$ with few shots, we 
assign random slopes only to a limited set of shot-types and treat the remaining shot-type effects as fixed effects. This design keeps the model additive on the log scale and stabilizes computation; together with sum-to-zero constraints that fix the overall baseline and regularizing hyperpriors (described below) that shrink the variances and temporal correlations of the seasonally correlated effects towards their base models, it yields an appropriate degree of partial pooling. Furthermore, the equicorrelation structure across seasons for team$\times$area and shot-type-specific effects allows us to capture both persistence and year-to-year variation, while the season-specific team$\times$side effects flexibly absorb coarse directional biases within each season. 
Note that imposing, for each season and area, a sum-to-zero constraint across teams on the season-correlated effects could yield more discriminative (i.e., more relative) team differences, but we did not pursue this due to implementation difficulties.

From a likelihood perspective, we retain a Poisson specification with a possession-based offset rather than switching to a negative-binomial likelihood with an additional dispersion parameter. In our setting, the dominant sources of overdispersion are heterogeneity across teams, regions, and shot types, which are already modeled through the random effects; adding a separate dispersion parameter at the likelihood level would effectively layer extra overdispersion on top of these components, complicating interpretation and identifiability. Prior predictive checks (Section~\ref{sec:prior_spec_ppc}) did not indicate substantial lack of fit at the aggregated cell level, supporting the adequacy of the Poisson formulation. For similar reasons, we do not impose additional spatial smoothing priors (e.g., conditional autoregressive structure) across neighboring regions. Instead, predefined court regions are treated as discrete areal categories
with independent random effects (i.e., we do not impose any spatial
smoothing across neighboring regions): this preserves sharp team-specific hot and cold spots, avoids blurring practically meaningful contrasts between regions, and keeps the latent Gaussian field sufficiently low-dimensional for efficient INLA-based inference.

\subsubsection{Priors for random effects and hyperparameters}

The random effects are modeled as
\begin{align}
  u_i &\stackrel{\mathrm{iid}}{\sim} 
    \mathcal{N}(0,\sigma^2_{\mathrm{team}}), \\
  v_a &\stackrel{\mathrm{iid}}{\sim} 
    \mathcal{N}(0,\sigma^2_{\mathrm{area}}), \\
  \mathbf{w}_{ia} &:= (w_{ia1},\ldots,w_{iaS})^\top 
      \sim \mathcal{N}\!\bigl(\mathbf{0},\,
        \sigma^2_{\mathrm{ta}}\,\mathbf{R}_S(\rho_{\mathrm{ta}})\bigr), \\
  z_{ids} &\stackrel{\mathrm{iid}}{\sim} 
      \mathcal{N}(0,\sigma^2_{\mathrm{ts}}), \\
  \mathbf{r}^{(m)}_{ia} &:= \bigl(r^{(m)}_{ia1},\ldots,r^{(m)}_{iaS}\bigr)^\top 
      \sim \mathcal{N}\!\bigl(\mathbf{0},\,
        \sigma^2_{m}\,\mathbf{R}_S(\rho_{m})\bigr),
      \quad m=1,\ldots,M,
\end{align}
where $\mathbf{R}_S(\rho)$ denotes the $S\times S$ equicorrelation matrix with ones on the diagonal and off-diagonal elements equal to $\rho$.

For the variance and correlation hyperparameters, we use PC priors. In particular, for given thresholds $U_{\bullet}, V_{\bullet}$ and 
tail probabilities $\alpha_{\mathrm{prec}}, \alpha_{\mathrm{cor}}$, we specify
\begin{align}
  \Pr(\sigma_{\bullet} > U_{\bullet}) &= \alpha_{\mathrm{prec}}, \qquad
  \Pr(|\rho_{\bullet}| > V_{\bullet}) = \alpha_{\mathrm{cor}},
\end{align}
so that large standard deviations and correlations close to $\pm 1$ are a priori penalized, while moderate values remain plausible. In all analyses we fix $\alpha_{\mathrm{prec}} = 0.05$, so that there is
only $5\%$ prior probability that any standard deviation exceeds its
scale parameter $U_{\bullet}$. For the season-correlation parameters
$\rho_{\mathrm{ta}}$ and $\rho_{m}$ we use PC-cor priors with
$V_{\bullet} = 0.7$ and $\alpha_{\mathrm{cor}} = 0.7$, that is,
\begin{align}
  \Pr\bigl(|\rho_{\bullet}| > 0.7\bigr) = 0.7,
\end{align}
which encodes a prior preference for moderate-to-strong correlations
while still shrinking towards $\rho_{\bullet}=0$.

The detailed calibration of other thresholds using prior predictive checks is described in 
Section~\ref{sec:prior_spec_ppc}.

\subsection{Prior specification and prior predictive checks}
\label{sec:prior_spec_ppc}
As described above, we place PC priors on the standard deviations and
seasonal correlation parameters of the random effects. In practice, we
tune the scale parameters $U_{\bullet}$ that control the typical size of
the standard deviations, and we divide them into two groups: a common
scale $U_{\mathrm{sd}}$ for the main random effects (team, area, and
team$\times$area) and a separate scale $U_{\mathrm{slope}}$ for the
shot-type-specific random slopes $r^{(m)}_{ias}$.

We follow a Bayesian workflow that emphasizes prior predictive checks to
calibrate these hyperparameters and detect implausible model behavior
\citep{gabry2019visualization}. Concretely, for each candidate's choice of
PC prior scales $(U_{\mathrm{sd}}, U_{\mathrm{slope}})$, we simulate
from the prior predictive distribution of the STAMP model, and compare
summaries of the simulated shot charts to simple empirical benchmarks
derived from the B.LEAGUE data, as described below.

In particular, to assess whether a given pair $(U_{\mathrm{sd}}, U_{\mathrm{slope}})$
produces realistic prior predictive behavior, we consider the summary
statistics of the replicated data. Let $T_{\mathrm{cell},95}$ denote the
$95$th percentile of the empirical cell-level rates
$Y_{isadk}/E_{is}$ across all cells $c = (i,s,a,d,k)$, and let $T_{\mathrm{tot},95}$
denote the $95$th percentile of the team-by-season total counts
$\sum_{a,d,k} Y_{isadk}$ over $(i,s)$. For each configuration of
$(U_{\mathrm{sd}}, U_{\mathrm{slope}})$ we generate $800$ prior predictive
replicates and compute two-sided $p$-values
\begin{align}
  p_{\mathrm{cell},95}
  \;=\;
  2\min\bigl\{ \Pr(T_{\mathrm{cell},95}^{\mathrm{rep}}
     \le T_{\mathrm{cell},95}^{\mathrm{obs}}),
               \Pr(T_{\mathrm{cell},95}^{\mathrm{rep}}
     \ge T_{\mathrm{cell},95}^{\mathrm{obs}})\bigr\},
\end{align}
and analogously $p_{\mathrm{tot},95}$ for $T_{\mathrm{tot},95}$. The
probabilities are approximated by empirical proportions over the 800
replicates. Following a conventional prior predictive check, we regard a
configuration as acceptable if both $p_{\mathrm{cell},95}$ and
$p_{\mathrm{tot},95}$ exceed a nominal threshold (here $0.05$).

We performed a grid search over
\begin{align}
  U_{\mathrm{sd}} \in \{0.5, 0.8, 1.0, 1.2, 1.5\},
  \qquad
  U_{\mathrm{slope}} \in \{0.5, 0.8, 1.0, 1.2, 1.5\},
\end{align}
resulting in $25$ candidate configurations. Among these, $7$ pairs
satisfied the prior predictive criterion
($p_{\mathrm{cell},95} > 0.05$ and $p_{\mathrm{tot},95} > 0.05$):
\begin{align}
  (U_{\mathrm{sd}}, U_{\mathrm{slope}}) \in
  \{(1.5,1.5), (1.5,1.2), (1.5,1.0), (1.5,0.8), (1.5,0.5),
    (1.2,1.2), (1.2,0.5)\}.
\end{align}
We regard these seven settings as admissible priors and use all of them
in a subsequent prior sensitivity analysis. For brevity, we report model-comparison results for a representative configuration from
this admissible set.

\subsection{Model fitting and evaluation}
\label{sec:model_fit_eval}

All models are fitted in a Bayesian framework using the latent Gaussian
structure described in Section~\ref{sec:proposed_model} and the PC
priors and prior predictive calibration in
Section~\ref{sec:prior_spec_ppc}. In what follows, we distinguish
between \emph{regular-season} possessions, which are used for model
fitting, and \emph{post-season} possessions, which are used only for
out-of-sample evaluation (see Section~\ref{sec:data_preproc} for details
of the dataset construction).

\subsubsection{Data split (fitting and evaluation)}
\label{sec:split}

We train all models on regular-season possessions and evaluate their
out-of-sample performance on post-season possessions, reflecting the
intended deployment scenario in which a model fitted on past seasons is
used to assess post-season tendencies. In total, the dataset contains
$306{,}444$ field-goal attempts, of which $296{,}768$ shots
($96.84\%$) occur in the regular season and $9{,}676$ shots
($3.16\%$) occur in the post-season. Regular-season shots contribute to
the likelihood used for parameter estimation, whereas post-season shots
are held out and used only for out-of-sample predictive evaluation.

\subsubsection{Model configurations}
\label{sec:comparison}

We compare a family of $16$ candidate specifications obtained by
switching on or off four groups of random effects. Concretely, we
introduce binary indicators
\(\texttt{use\_ta}, \texttt{use\_ts}, \texttt{use\_jump}, \texttt{use\_step}
\in \{0,1\}\) controlling the inclusion of
team$\times$area effects $w_{ias}$ (\texttt{use\_ta}),
team$\times$side effects $z_{ids}$ (\texttt{use\_ts}), and
shot-type-specific random slopes for jump shots
(\texttt{use\_jump}) and step-back / pull-up jump shots
(\texttt{use\_step}). We then consider all combinations
\begin{align}
  \{\texttt{use\_ta}, \texttt{use\_ts}, \texttt{use\_jump}, \texttt{use\_step}\}
  \in \{0,1\}^4,
\end{align}
so that the most complex specification uses all four components and the
simplest one includes only the fixed effects and the team- and
area-level main effects.

Random slopes are introduced only for the \texttt{jump\_shot} and
\texttt{step\_pull} categories. These shot-types have sufficiently many
attempts per team to support additional parameters, and they exhibit
substantially larger spatial variability than other categories, making
them natural candidates for capturing team-specific spatial tendencies.
For the remaining shot-types we retain only fixed effects, which helps
avoid over-parameterization and instability in cells $c$ with
few observations. All candidate models share the same fixed-effect
structure and the same family of hyperpriors; only the presence or
absence of these random-effect blocks differs.

\subsubsection{Fitting via INLA}
\label{sec:fit_inla}

For each model configuration, we aggregate the play-by-play data to the
cell level $c$ and form regular-season counts
$Y^{\mathrm{reg}}_c$, post-season counts $Y^{\mathrm{post}}_c$, and
team-by-season exposure terms $E^{\mathrm{reg}}_{is}$ and
$E^{\mathrm{post}}_{is}$. The model in \eqref{eq:rs-general} is then
fitted to the regular-season data by treating $Y^{\mathrm{reg}}_c$ as
the response and $\log E^{\mathrm{reg}}_{is}$ as the offset. We use a
Poisson likelihood with a log link and exploit the latent Gaussian
structure to perform approximate Bayesian inference via INLA, as implemented in the
\texttt{R-INLA} package. For each fit we request posterior summaries for
all fixed and random effects and for the linear predictors $\eta_c$, and
enable configuration sampling (\texttt{config = TRUE}) so that we can
draw posterior samples from the latent Gaussian field using
\texttt{inla.posterior.sample}. These posterior draws are later used to
evaluate out-of-sample predictive performance on the post-season data via
Monte Carlo integration of the log predictive density.

All random-effect blocks are implemented as independent Gaussian effects
with \texttt{model="iid"} in \texttt{R-INLA}. Seasonally correlated
team$\times$area effects $w_{ias}$ and shot-type-specific slopes
$r^{(m)}_{ias}$ are represented using the \texttt{group} mechanism with
\texttt{group = season} and an exchangeable correlation structure across
seasons, corresponding to the equicorrelation matrix
$\mathbf{R}_{S}(\rho)$ in Section~\ref{sec:proposed_model}. The
team$\times$side effects $z_{ids}$ are modeled as season-specific i.i.d.\
effects without an explicit temporal correlation, with zero-sum
constraints over $d$ for each $(i,s)$ imposed via the \texttt{extraconstr} argument. To improve numerical stability we add a small ridge term to the fixed effects (through \texttt{control.fixed}) and use \texttt{strategy="adaptive"} and \texttt{int.strategy="ccd"} for the INLA internal approximations.

\subsubsection{Criterion on fitting (regular-season) data}
On the regular-season data used for fitting, we monitor the log
pseudo-marginal likelihood (LPML; \cite{geisser1979predictive,held2009posterior}). LPML is defined as
\begin{align}
  \mathrm{LPML}_{\mathrm{reg}}
  \;=\;
  \sum_{c}
    \log\bigl\{\mathrm{CPO}_c\bigr\},
\end{align}
where $\mathrm{CPO}_c$ denotes the conditional predictive ordinate for
cell $c$ (regular-season observation), as returned by INLA. Larger
values of $\mathrm{LPML}_{\mathrm{reg}}$ correspond to better in-sample
predictive fit. In the result tables we report $\mathrm{LPML}_{\mathrm{reg}}$ as a secondary diagnostic to check that the models selected by the post-season criterion are also reasonable on the fitting data.

\subsubsection{Out-of-sample evaluation on the post-season}
Our primary model-comparison criterion is the expected log predictive density on the post-season data. For each fitted model we keep the posterior obtained from regular-season data and evaluate predictive performance on the held-out post-season counts
$Y^{\mathrm{post}}_{c}$ with exposure $E^{\mathrm{post}}_{is}$. Let $\theta$ denote the collection of fixed and random effects and hyperparameters; for each cell $c$ with $E^{\mathrm{post}}_{is}>0$ we consider the log predictive density
\begin{align}
  \log p\bigl(Y^{\mathrm{post}}_{c} \mid \theta\bigr),
\end{align}
where the Poisson mean is
$\mu^{\mathrm{post}}_{c} = E^{\mathrm{post}}_{is} \exp(\eta_{c})$.

From each INLA fit, we draw $J$ posterior samples
$\theta^{(1)},\dots,\theta^{(J)}$ of the latent Gaussian field (here
$J=400$), and approximate the pointwise log predictive densities by
Monte Carlo,
\begin{align}
  \widehat{\ell}_c
  \;=\;
  \log \Biggl(
    \frac{1}{J} \sum_{j=1}^{J}
      p\bigl(Y^{\mathrm{post}}_{c} \mid \theta^{(j)}\bigr)
  \Biggr).
\end{align}
Summing over all post-season cells yields the Monte Carlo estimate of the
expected log predictive density,

\begin{align}
    \widehat{\mathrm{elpd}}_{\mathrm{post}}^{\mathrm{MC}}
  \;=\;
  \sum_{c}
    \widehat{\ell}_c.
\end{align}
We use $\widehat{\mathrm{elpd}}_{\mathrm{post}}^{\mathrm{MC}}$ as the primary selection criterion, with larger values indicating better out-of-sample predictive performance on post-season shot charts. In the result tables we report $\widehat{\mathrm{elpd}}_{\mathrm{post}}^{\mathrm{MC}}$ together with $\mathrm{LPML}_{\mathrm{reg}}$ to jointly assess extrapolative performance and fit to the regular-season data.

All computations are carried out in \texttt{R} using a single BLAS thread for reproducibility. With the present dataset (on the order of $3\times 10^5$ shots aggregated to team-season-area-side-shot-type cells), each INLA fit completes within roughly a minute on a standard workstation-class CPU.

\subsubsection{Additional team-level shot-type slopes}

In addition to the main family of 16 candidate models based on
team$\times$area and shot-type-specific slopes for jump and
step-back/pull-up shots (Section~\ref{sec:model_fit_eval}), we also
considered an extended set of models that introduce team-level
shot-type slopes for the remaining categories (\texttt{lay\_up}, \texttt{floater},
\texttt{rim\_finishes}, \texttt{fade\_turn}). Concretely, we kept the team$\times$area
slopes for \texttt{jump\_shot} and \texttt{step\_pull} unchanged and switched on/off the
presence of team-level slopes for each of the four additional shot types, combined with \texttt{use\_ts} $\in\{0,1\}$, yielding $2\times 2^4 = 32$ additional candidates. The prior hyperparameters were fixed to the values calibrated in Section~\ref{sec:prior_spec_ppc}, with $U_{\mathrm{sd}}=1.5$,
$U_{\mathrm{slope}}=1.0$, $\alpha_{\mathrm{prec}}=0.05$ and
$V=0.7$, $\alpha_{\mathrm{cor}}=0.7$. We evaluated these extended models in the same way as the main candidates, using
$\widehat{\mathrm{elpd}}_{\text{post}}^{\mathrm{MC}}$ on post-season data and $\mathrm{LPML}_{\text{reg}}$ on regular-season data.

\backmatter

\section*{Competing interests}
The authors have no competing interests to declare that are relevant to the content of this article.

\section*{Author Contributions}
K.Y. contributed to the study conception and design. Data preparation, modeling, and analysis were performed by K.Y. The first draft of the manuscript was contributed by K.Y. and K.F. All authors read and approved the final manuscript.

\section*{Data availability}
The data of the research is obtained by participating in a competition hosted by the academic organizations.
The central idea of this study was independent of the competition (not restricted by the competition).
Data acquisition was based on the contract between the basketball league and Data Stadium, Inc., but not between the players/teams and us. 

\section*{Funding Information}
K.~Yamada gratefully acknowledges support from the “THERS Make New Standards Program for the Next Generation Researchers”.
This work was financially supported by JSPS KAKENHI Grant Number 23H03282.

\ifarxiv
\bmhead{Acknowledgments}
The data used in this study were provided by the Institute of Statistical Mathematics, Academy for Statistics and Data Science, Research Organization of Information and Systems, and Data Stadium Inc. 
\fi

\begin{appendices}
\setcounter{table}{5}
\setcounter{figure}{6}
\setcounter{equation}{18}

\end{appendices}

\newpage
\bibliography{sn-article}

\end{document}